\documentclass[runningheads]{llncs}
\usepackage{graphicx}
\usepackage{rotating}
\usepackage{lscape}
\usepackage{amsfonts}
\usepackage{algorithm}
\usepackage{algorithmicx}
\usepackage{algpseudocode}
\usepackage{amsmath}
\usepackage{multirow}
\usepackage{array}
\usepackage{subfigure}
\usepackage{enumerate}
\usepackage{amssymb}

\begin{document}
\title{Community-Based Service Ecosystem Evolution Analysis}
\titlerunning{Abbreviated paper title}
\author{Mingyi Liu \and
Zhiying Tu \and
Xiaofei Xu \and
Zhongjie Wang}
\authorrunning{M. Liu et al.}
\institute{School of Computer Science and Technology, Harbin Institute of Technology, China \\
\email{\{liumy, tzy\_hit, xiaofei, rainy\}@hit.edu.cn}}
\maketitle              % typeset the header of the contribution

\begin{abstract}
%% 避免关于商业决策的过多重复， 避免对技术类别的服务生态系统演化的直接批评
The prosperity of services and the frequent interaction between services contribute to the formation of the service ecosystem. Service ecosystem is a complex dynamic system with continuous evolution. Service providers voluntarily or compulsorily participate in this evolutionary process and face great opportunities and challenges. Existing studies on service ecosystem evolution are more about facilitating programmers to use services and have achieved remarkable results. However, the exploration of service ecosystem evolution from the business level is still insufficient. To make up this deficiency, in this paper, we present a method for analyzing service ecosystem evolution patterns from the perspective of the service community. Firstly, we train a service community evolution prediction model based on the community evolution sequences. Secondly, we explain the prediction model, showing how different factors affect the evolution of the service community. Finally, using the interpretable predictions and prior knowledge, we present how to assist service providers in making business decisions. Experiments on real-world data show that this work can indeed provide business-level insights into service ecosystem evolution. Additionally, all the data and well-documented code used in this paper have been fully open source.

\end{abstract}
\section{Instruction}

Along with the trend of Everything as a Service (EaaS), services flourish drastically both on the Internet and in the real world. These services have become much more interconnected and have formed various service ecosystems, such as ``Internet of Services'', ``Smart Planet"\footnote{https://www.ibm.com/smarterplanet/us/en/} and ``Big Services''\cite{xu2015big}. Nowadays, the service ecosystem continues to evolve every day, with the emergence, prosperity, integration, and decline of a large number of service individuals/service populations. The service providers in the service ecosystem are actively or passively involved in this evolution process and faced with great opportunities and challenges. It is of great significance to explore the evolution patterns and predict the evolution trend of the service ecosystem based on historical information, which can assist service providers in making business decisions to survive better in the competitive environment.

%% 参考文献得换商业相关的==> (现有研究的缺失)
%% 先写服务生态系统演化研究可以分为技术层面研究和商业层面的研究，目前已有的研究主要是关注技术层的研究，这个带来了balabala的好处。但是现实中，在技术之上还有商业研究，商业层面的探究可以带来更大的利益。
The existing studies on service ecosystem evolution mainly focus on solving the traditional service computing problems, such as service recommendation\cite{10.1109/tase.2013.2297026,10.1007/978-3-662-45391-9_43,10.1109/icws.2014.17} and service discovery\cite{10.1007/978-3-030-03596-9_44,wan2016incorporating}, by exploring the changes in the overall network propriety of the service ecosystem or the changes of the individual attributes of the service. 
%% 肯定价值
These studies mainly focus on technical-level service and  have yielded many commendable results, helping programmers to develop services faster and better. 
%% However, 这个来肯定在技术层面的价值，但是现实中还有更宽泛，业务层面也可以做分析，可以带来更大的价值。
In the real world, in addition to technical-level services, there are more common business-level service. Exploring the evolution of business-level service ecosystems can bring significant benefits and unique insights that cannot be provided by exploring technical-level service ecosystems, such as as service market trend development prediction and business decision assistance.
However, there is still a lack of studies on the evolution of business-level service ecosystem is still insufficient, which is mainly caused by the lack of data and corresponding methods.

% Some existing studies involve the analysis of patterns in the service ecosystem. These studies mainly analyze some static service community patterns in the service ecosystem, such as service communities with different functional topics\cite{10.1109/scc.2017.23,yang2017fast}, which ignore the dynamic evolution of the service ecosystem.

In this paper, we explore the evolution patterns and laws of the business-level service ecosystem from the perspective of the service community, to provide helpful insights on the evolving service ecosystem and suggestions for business decisions. To achieve this, we train and explain a service community evolution event prediction model. 
%% MSEM 这个加一个定于从句解释 (解释了一下从哪来，用这个的原因？解决上面说的没数据的问题)
Specifically, we first extract a subgraph\footnote{In the remainder of this paper, we will treat this subgraph as a service ecosystem.} of historical interaction between stakeholders and services from Multilayer network-based Service Ecosystem Model (MSEM)\cite{liu2020datadriven}, which is constructed from the social media news data and solves the data deficiency of the business-level service ecosystem. Next, a series of service ecosystem snapshots at different times are generated, and the static community detection algorithm is applied to these snapshots to obtain the service community structure at different times. Then, by using the Group Evolution Detection (GED) algorithm\cite{10.1007/s13278-012-0058-8}, the service community structures at adjacent moments are aligned and the community evolution events are identified, including \textit{forming}, \textit{continuing}, \textit{growing}, \textit{shrinking}, \textit{splitting}, \textit{merging}, and \textit{dissolving} \cite{palla2007quantifying}.  After that, sequences are created to describe the evolution of the service community. Each sequence is consists of several preceding service community features and the corresponding evolution event, which serve as input for the prediction model. Finally, the service community evolution event prediction model is trained, and SHapley Additive exPlanations (SHAP)\cite{NIPS2017_7062} is used to explain this machine learning model. Based on the explainable result, we summarized some patterns of service community evolution in the evolving service ecosystem and show how this explainable result can be contributed to assisting business decisions.

The main contributions of this paper are as follows:
\begin{enumerate}[1)]
    \item Different from traditional research on service ecosystem evolution, we explore the evolution patterns of the service ecosystem from the perspective of the service community. To the best of our knowledge, this is the first time that community evolution analysis is applied to the service ecosystem.
    %%避免简单重复(1.分析模式发现规律，2. 进行预测，进行决策)

    \item A service community evolution prediction model is trained and explained. Applying the appropriate visualization to the interpretation results, we summarize some evolution patterns of service communities in the evolving service ecosystem.
    \item We used the trained prediction model to predict the evolution trend of a service community and show how this prediction can help service providers make business decisions.
    
\end{enumerate}

 All the data and algorithms used in the experiments have been fully open-sourced\footnote{https://github.com/icecity96/icsoc-2020}, to help researchers in this domain for more deep research.
 
The remainder of this paper is organized as follows. Section \ref{sec:related_work} introduces related work. Section \ref{sec:framework} gives the process and steps of exploring service ecosystem evolution from the perspective of the service community. Section \ref{sec:discuss} evaluates our approach on real-world data and presents a case study in the bike-sharing. And the last section is the conclusion.

\section{Related Work}\label{sec:related_work}
\subsection{Service Evolution \& Service Ecosystem Evolution}
There have been a number of studies investigating the evolutionary properties of service and service ecosystems. These evolutionary studies focus on the state changes of service individuals and the changes of service network topology to help programmers select appropriate service for integration into their applications. In other words, these studies are mainly aimed at solving the classic problems in service computing, such as service recommendation, service discovery, and service deployment.

For example, Adalberto et al\cite{8094458} aggregated structural, deployment, and runtime information of an evolving microservice system in one model, which provides actionable insights to help developers manage service upgrades, architectural evolution, and changing deployment trade-offs. Fokaefs and Stroulia\cite{fokaefs2014wsdarwin} explored service evolution by comparing changes on interface description of a service. Huang et al\cite{10.1109/tase.2013.2297026} used a network link prediction method to study both usage patterns and the evolution traces of entire \textit{ProgrammableWeb} ecosystem for service recommendation. Adeleye et al\cite{10.1007/978-3-030-03596-9_44} organized web API services into a complex network graph by considering relations between them and studied the evolution of services for service discovery by analyzing the changes of this complex network. Yang et al\cite{10.1109/icws.2014.17}  used Latent Dirichlet Allocation (LDA) and time series prediction to extract service evolution patterns for time-aware service recommendation.

These studies are all on technical-level service ecosystem, and the have achieved respectable results. In our view, in addition to the technical-level service ecosystem, the business-level ecosystem is a much broader service ecosystem in real world, and exploring the evolution of the business-level service ecosystem can bring more unique insights and significant benefits.

\subsection{Community Evolution Prediction}
As for the research on community evolution prediction, researchers mainly focus on (1) how to make prediction results more accurate and (2) explore the influence of different factors on prediction results. For example, Saganowski et al\cite{saganowski2015predicting} presented and analyzed two methods to predict the near future of the community. They summarized that: (1) the longer community history cause higher quality of prediction, and (2) the most recent history of the community has the most influence on its next change. Dakiche et al\cite{10.1145/3297280.3297484} improved the accuracy of prediction by using change rates of features that describe a community throughout its evolution life-cycle rather than absolute values of features and achieved excellent results on DBLP and Facebook datasets. Dakiche et al\cite{8491668} also explored the effect of timeframe type and size on community evolution prediction on Facebook and Higgs Twitter datasets.

Inspired by the success of community evolution prediction in social network analysis, we applied it to the analysis of service ecosystem evolution. The experiment shows that there are some similar patterns in the evolution of the service community and social community and some unique patterns in the evolution of the service community.

\section{The Framework of Service Community Evolution Analysis and Prediction\vspace{-1em}}\label{sec:framework}

\begin{figure}
    \centering
    \includegraphics[width=\textwidth]{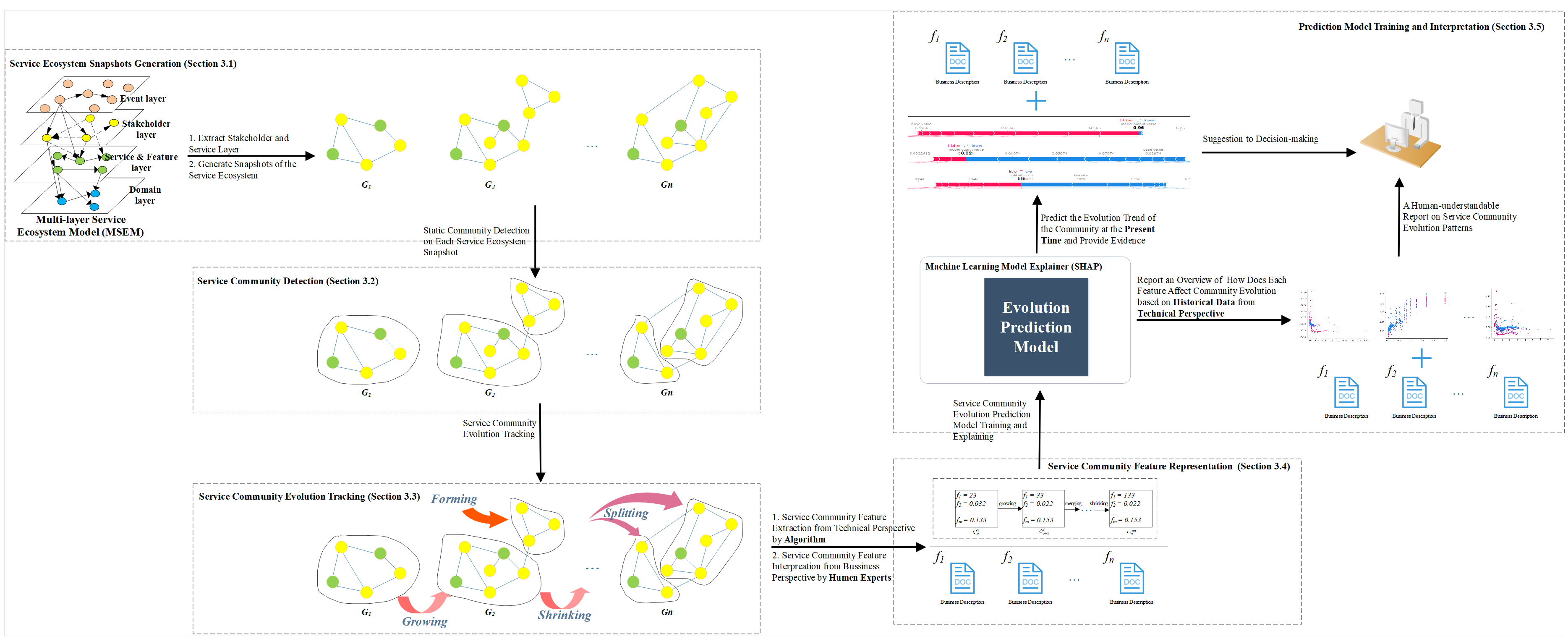}
    \caption{The framework of service community evolution analysis and prediction}
    \label{fig:framework}
        \vspace{-1em}

\end{figure}

Fig.~\ref{fig:framework} shows the framework of service community evolution analyzing and prediction in an evolving service ecosystem. The five parts of the framework are discussed in detail in the next five subsections of this section.

\subsection{Service Ecosystem Snapshots Generation}
We extract the stakeholder layer and service \& feature layer involved in evolution from MSEM as the evolving service ecosystem studied in this paper. This ecosystem can be represented as $G=(V, E)$, where $V$ represents the stakeholder and service in the ecosystem, which are collectively referred to $entity$ for convenience. $E = \{(u, v, r, t) | u, v \in V \}$ refers to the historical interactions between entities, where $r$ is the type of interaction, such as conflict and cooperation, and $t$ is the time of interaction.

In order to analyze the evolution of service ecosystem, it is segmented into $n$ adjacent snapshots $\mathbb{G} = (G_0, G_1, \dots, G_n)$. Each snapshot $G_i = (V_i, E_i)$ represents a \textbf{weighted undirected} graph with only the set of entities and edges that can be observed at time $T_i$. $E_i = \{(u, v, w) | u, v \in V_i, w \in \mathbb{R}^{+}\}$ represents the relation between entities with $w$ indicates the tightness of the relation. The main challenge in generating service ecosystem snapshots is how to determine the weight of each edge in the snapshot. We design Algorithm \ref{alg:snapshot} to generate service ecosystem snapshots based on the following assumptions:
\begin{enumerate}[1)]
    \item The tightness of relation between entities is the cumulative result of their historical interactions.
    %% 让人误解只有正负号
    \item Different types of interactions have different impacts, which means that the impact of interactions on the weight is different, and the impact can be either positive or negative.
    \item The influence mechanism of different types of interaction on weight is different. In this paper, these mechanisms are summarized, including the \textbf{stability mechanism} that is not affected by time, the \textbf{aging mechanism} that declines with time and the \textbf{mutation mechanism} that is not affected by historical interaction.
\end{enumerate}

\begin{algorithm}
\caption{Generation Service Ecosystem Snapshot} \label{alg:snapshot}
\begin{algorithmic}[1]
\Require Service ecosystem $G=(V, E)$, timestamp $T_i$, interactions impacts $I$, interactions mechanism $M$.
\Ensure Service ecosystem snapshot $G_i$ at time $T_i$.
\State initial $G_i$ with all weight $w$ are $0$.
\State {$edges \gets E$ sorted by $t$ in ascending order}
\For {$(u, v, r, t)$ \textbf{in} $edges$}
    \If {$t > T_i$}
        \State \textbf{break}
    \EndIf
    \If {$M[r]$ \textbf{is} stability}
        \State {$w_{u, v} \gets w_{u, v} + I[r]$}
    \ElsIf {$M[r]$ \textbf{is} mutation}
        \State {$w_{u, v} \gets I[r]$}
    \ElsIf{$M[r]$ \textbf{is} aging}
        \State {$w_{u,v} \gets w_{u, v} + I[r] \times aging\_coeff(T_i, t)$}
    \EndIf
\EndFor
\State {Remove the isolated nodes and edges with non-positive weights in $G_i$}
\State \Return{$G_i$}
\end{algorithmic}
\end{algorithm}
The aging coefficient used in Algorithm \ref{alg:snapshot} can be calculated as follows:
\begin{equation}
    aging\_coeff(t_i, t_j) = \left\{ 
    \begin{aligned}
        & \frac{1}{\max (\frac{t_i - t_j}{a}, 1)}, & t_i - t_j < b \\
        & 0, & \text{otherwise}
    \end{aligned}
    \right.
\end{equation}
%% 详细一点的解释
$aging\_coeff(\centerdot)$ is a linear decay function, where $a$ and $b$ are two parameters used to indicate the period of decline and the maximum duration, which are set to 30 days and 365 days respectively in this paper.

\subsection{Service Community Detection}
Community detection is a popular research topic in the field of social network analysis. Several methods have been summarised in \cite{khan2017network}. Each detection method has its characteristics and applicable scenarios, and the Louvain algorithm\cite{blondel2008fast} with the advantages of easy implementation, no specific parameters, fast and good partition quality is used in this paper. It should be noted that one could use any other static community detection algorithm.

Due to the instability of the community detection, in order to obtain stable communities, we only keep the community of size more than $3$. The $k$ communities detected in the $G_i$ are denoted by $C_i = (C_i^1, C_i^2, \dots, C_i^k)$, where each $C_i^p \in C_i$ is also a graph denoted by $(V_i^p, E_i^p)$. Key nodes (leaders), which have higher social position than surrounding nodes, is also detected in each community and denoted by $K_i^p = \{v_i^1, v_i^2, \dots, v_i^m\}$. The social position $SP$ is represented by the PageRank score in this paper.

\subsection{Service Community Evolution Tracking} \label{sec:tracking}
%% 这部分没有展开解释，本质上是别人的方法，使用I值来判断演化事件。
%% 这部分写法上是参考 Community Evolution Prediction in Dynamic Social Networks Using Community Features’ Change Rates
GED\cite{10.1007/s13278-012-0058-8} is applied as a community evolution tracker that can detect evolution events through the community structure of consecutive snapshots, including \textit{forming}, \textit{continuing}, \textit{growing}, \textit{shrinking}, \textit{splitting}, \textit{merging}, and \textit{dissolving} \cite{palla2007quantifying}. The most important component in GED is a structural method called \textit{inclusion}, which allows to evaluate the inclusion of one community in another. It takes into account both quantity and quality:
\begin{equation}
    I(C_t^i, C_{t+1}^j) = \frac{\overbrace{|C_t^i \cap C_{t+1}^j|}^{\text{community quantity}}}{C_t^i} \frac{\sum_{x \in C_t^i \cap C_{t+1}^j}SP_{C_t^i}(x)}{\underbrace{\sum\nolimits_{x \in C_t^i} SP_{C_t^i}(x)}_{\text{community quality}}}
\end{equation}
where $SP_{C_t^i}$ is the social position of $x$ in community $C_t^i$.

$\alpha$ and $\beta$ are the GED method parameters, which are used to adjust the method to the particular network and community detection method. In this paper, they are both set to $0.5$.

\subsection{Service Community Features Representation}\label{sec:features}
Based on the scale of our data and the demand for interpretability, we manually selected features rather than using representation learning to acquire features. Community structural features and community key nodes features are selected based on the following intuition:
\begin{enumerate}[1)]
    \item The service community is a dense subgraph composed of closely related entities, so the aggregated topology features in the community are very important.
    \item Key nodes, which promote the formation of the community, have more influence than other members of the community and usually lead the development of the community. We added the features of the key nodes to our consideration.
\end{enumerate}
Table \ref{tab:features} summarizes the selected features as well as formulations and business explanations.

\begin{table}[htbp]
\caption{Selected service community features as well as formulations and explanations}\label{tab:features}
\begin{tabular}{|p{0.15\textwidth}|p{0.325\textwidth}|p{0.2\textwidth}<{\centering}|p{0.325\textwidth}|}
\hline
Feature                                                      & Description                                                                                                        & \multicolumn{1}{l|}{Formulation}                                                                                                                                                 & Business explanation                                                                                                      \\ \hline
size                                                         & Number of node within the community.                                                                               & $|V_t^i|$                                                                                                                                                   & Reflects the scale of the service community                                                                               \\ \hline
density                                                      & The portion of potential connections in a community that are actual connections                                    & $\frac{2|E_t^i|}{|V_t^i|(|V_t^i - 1|)}$                                                                                                                     & Reflects the degree of familiarity between the entities within the service community                                      \\ \hline
clustering                                                   & The average clustering coefficient $cc_{t}(x)$ of nodes in the community                                           & $\frac{\sum\nolimits_{x \in V_t^i}cc_{t}(x)}{|V_t^i|}$                                                                                                      & Reflects the degree of solidarity among the entities in the service community.                                            \\ \hline
\begin{tabular}[c]{@{}l@{}}average\\ closeness\end{tabular}  & The average closeness $cl_t(x)$ of nodes in the community                                                          & $\frac{\sum\nolimits_{x \in V_t^i}cl_{t}(x)}{|V_t^i|}$                                                                                                      & Reflects the importance of the service community to the service ecosystem.                                                \\ \hline
degree                                                       & The average \textbf{weighted} degree $wd_t(x)$ of nodes in the community                                           & $\frac{\sum\nolimits_{x \in V_t^i}wd_{t}(x)}{|V_t^i|}$                                                                                                      & Reflects the strength of internal and external interaction ability of the service community.                              \\ \hline
leadership                                                   & The centralization of the community                                                                                & $\frac{\max(\{wd_{t}(x) | x \in V_t^i \})}{(|V_t^i|-1)(|V_t^i|-2)}$                                                                                         & Reflects the single entity's maximum control over the entire service ecosystem.                                           \\ \hline
cohesion                                                     & The characterising strength of connections inside community in relation to connections outside                     & $\frac{\frac{\sum_{p \in C_t^i}\sum_{q \in C_t^i}w(p,q)}{(|V_t^i|-1)|V_t^i|}}{\frac{\sum_{p \in C_t^i}\sum_{q \not\in C_t^i}w(p,q)}{|V_t|(|V_t|-|V_t^i|)}}$ & Reflects the degree of solidarity within the service community.                                                           \\ \hline
\#KeyNodes                                                   & Number of key nodes in the community                                                                               & $|K_t^i|$                                                                                                                                                   & Reflects the number of leaders in the service community.                                                                  \\ \hline
activity                                                     & The max/sum/average weight of edges in the community (only max formulation is given)                               & $\max(\{w(p, q) | p \in C_t^i, q \in C_t^i\})$                                                                                                              & Reflects the activity of the service community, and also reflects the degree to which the service community is concerned. \\ \hline
\%NodeType                                                   & The ratio of different type nodes in the community (only the formulation for service nodes $V_{service}$ is given) & $\frac{\{v | v \in V_t^i \cap V_{service} \}}{|V_t^i|}$                                                                                                       & Reflects the proportion of stakeholders and services in the service community.                                            \\ \hline
Kdegree                                                      & The average \textbf{weighted} degree $wd_t(x)$ of key nodes in the community                                       & $\frac{\sum\nolimits_{x \in K_t^i}wd_{t}(x)}{|K_t^i|}$                                                                                                      & Reflects the strength of internal and external interaction ability of the key nodes in the service community.             \\ \hline
\begin{tabular}[c]{@{}l@{}}Kaverage\\ closeness\end{tabular} & The average closeness $cl_t(x)$ of key nodes in the community                                                      & $\frac{\sum\nolimits_{x \in K_t^i}cl_{t}(x)}{|K_t^i|}$                                                                                                      & Reflects the importance of the core members of the service community in the whole ecosystem                               \\ \hline
\end{tabular}
\end{table}

Using the features given in Table \ref{tab:features}, each service community $C_t^i$ can be denoted as vector $F_t^i=[f_{t,i}^1, f_{t,i}^2, \dots, f_{t,i}^{15}]$, where $f_{t, i}^j$ corresponds to the features listed in the table.
It is important to point out that some of the features have similar business explanations, but are reflected from different perspectives.

\subsection{Prediction Model Training and Interpretation}
After the steps mentioned in Section \ref{sec:tracking} and Section \ref{sec:features}, we can obtain the evolution sequence of the service community and the their vector representation at different times. To balance the amount of training data and the information contained in the data, we use the state of the current moment and the state of the two past moments to predict the evolution, so the sequence length is $3$, and the input can be denoted as follow:
\begin{equation}
X= F_{t-2}^i \oplus F_{t-1}^i \oplus F_{t}^i
\end{equation}
The prediction target is the evolutionary event mentioned in Section \ref{sec:tracking}, except \textit{forming}, because \textit{forming} has no past instances.

We use a random forest as prediction model based on the following reasons:
\begin{enumerate}
    \item Compared with several other prediction models, the random forest performs better on our real-world dataset. Detailed comparison is shown in Section \ref{sec:discuss}.
    \item Tree models do not require large amounts of data for training and are easy to interpret and understand.
\end{enumerate}
After the prediction model is trained. SHAP, which is a machine learning interpreter based on game-theoretic, are applied to explain the output of the prediction model. SHAP use \textit{shap values} to explain how different features affect the predicted results. Combining the shap values with the appropriate visualization, we create results that are easily understood by humans.

%% 这里要加强
The interpretation of the prediction model mainly serves two purposes: (1) Summarize the evolution patterns of service community based on historical data (train data). (2) Predict the evolution trend of a service community at present (test data) and provide evidence to assist in business decisions.

\section{Case Study} \label{sec:discuss}
In this section, we provide an case study on the bike-sharing service ecosystem. Since the bike-sharing industry has not been around for a long time, the Internet media has recorded a relatively complete development history, which provides us with an ideal data. Moreover, the bike-sharing service ecosystem has gone through a complete life circle, including growth, prosperity and decline. 

We extracted data on the bike-sharing filed before December 24, 2019 from the auto-built MSEM. Starting from August 1, 2016, a snapshot of the service ecosystem will be generated every 30 days.

\subsection{Snapshots Generation \& Service Community Detection}
\begin{figure}[!hbtp]
    \centering
    \includegraphics[width=\textwidth]{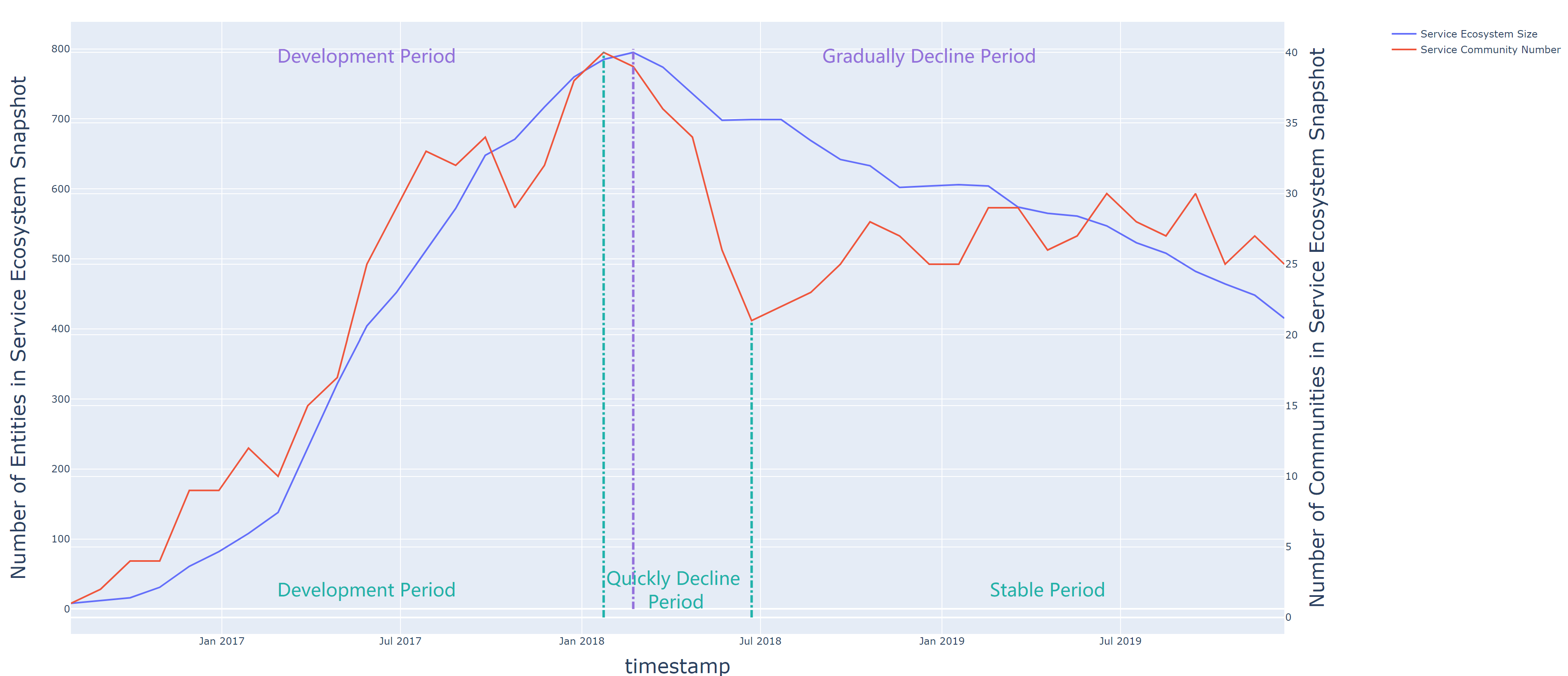}
    \caption{The evolution of the size of service ecosystems and the number of service communities}
    \label{fig:snapshots_size_community}
        \vspace{-1.5em}

\end{figure}

Fig. \ref{fig:snapshots_size_community} shows the size of service ecosystem (the blue line) and the number of service communities (the orange line) in the service ecosystem snapshot at different times. It can be seen that the size of service ecosystem and the number of service community both grew rapidly and reach their peak. However, the number of service communities is earlier the size of service ecosystem at both the time of high-speed growth and reaching the peak. This reflects that the service community is more sensitive to the evolution trend of the service ecosystem than the service ecosystem itself, and can perceive the evolution trend of the service ecosystem earlier. Decline comes after prosperity. Decline comes after a short period of prosperity. However, the decline mechanism of the service ecosystem scale is different from that of the number of service communities. The former is a gradual and slow decline, while the latter is a rapid decline after and then into a stable stage. Combining with the domain knowledge, we can know that the reduction of service ecosystem scale is due to the saturation of demand, and the change in the number of service communities is due to the fierce market competition.

To sum up, Fig. \ref{fig:snapshots_size_community} confirms the point we made in the introduction that service communities can indeed provide unique insights that service ecosystem as a whole cannot.

\subsection{Results of Service Evolution Tracking}
\begin{figure}
    \centering
    \includegraphics[width=\textwidth]{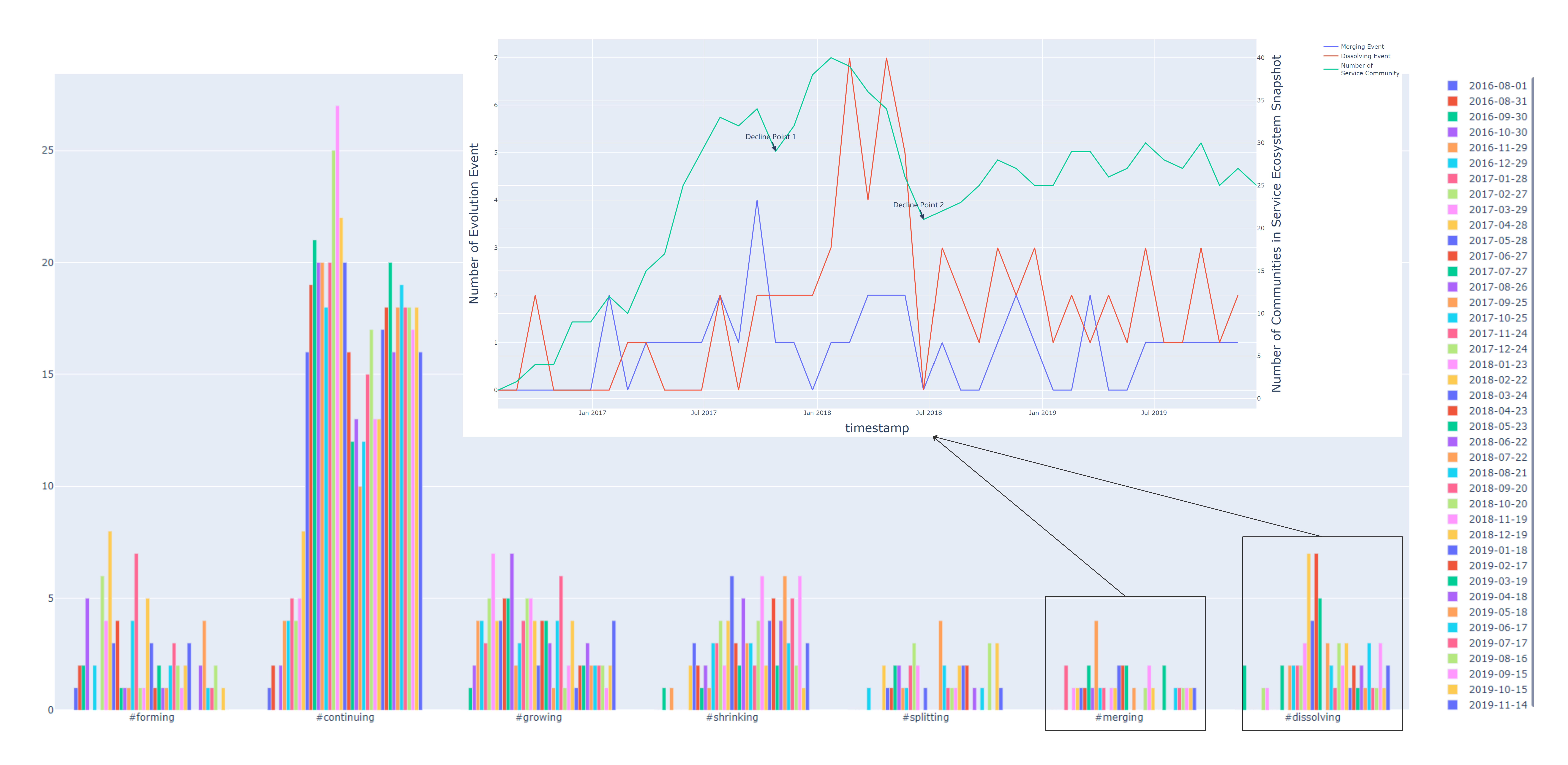}
    \caption{Distribution of service community evolution events}
    \label{fig:event_distribution}
        \vspace{-1.5em}

\end{figure}

Fig. \ref{fig:event_distribution} shows the distribution of service community evolution events over different time periods. $continuing$ is the most frequent evolution event. The formation and growth of service communities are mainly before 2018. A large number of communities died in mid-2018. Since late 2018, more service communities are starting to shrinking, which means that the competition in the bike-sharing market is coming to an end and the whole market is already saturated. Compared with other types of evolutionary events, service community merging and splitting events are significantly less.

By combining the distribution of service community evolution events and the change of the number of service communities, we can further understand the reasons for the change of the number of service communities. For example, it can be observed that two rapid decline of the number of service communities (labeled as \textit{decline point} in Fig. \ref{fig:event_distribution}) were mainly caused by $merging$ and $dissolving$ respectively. Furthermore, we can find that the subsequent effects of decline caused by the two type events are also different. The decline caused by $merging$ soon recovered and grew and flourish again, while the decline caused by $dissolving$ did not recover.

\subsection{Evaluate Service Community Evolution Prediction Model}
\begin{figure}
    \centering
    \includegraphics[width=\textwidth]{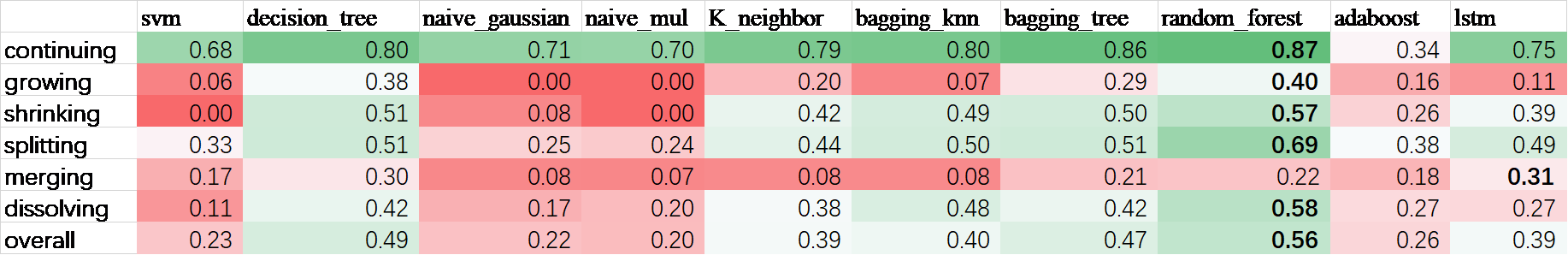}
    \caption{The $F1$ values of different prediction models}
    \label{fig:f1}
        \vspace{-1.5em}

\end{figure}

We divide the dataset containing $831$ samples into a train set and a test set, containing 665 samples and 166 samples, respectively. Fig. \ref{fig:f1} compares the $F1$ values of different prediction models on each evolution event on the test set.

As we can see, compared withe other models, the tree models (decision tree, bagging tree, random forest) performs well on our small dataset. The random forest get the highest $F1$ value for all evolution events except for $merging$. Deep learning method (LSTM) do not perform well, because of the small data size. All models have poor performance for \textit{merging}.

It should be pointed out that the focus of paper is not to find an optimal service community evolution prediction model. These models may vary greatly in different domains and datasets of different sizes. In this paper, we selected random forest for subsequent analysis because it performed best on our dataset.

\subsection{Evolution Patterns of Service Community}\label{sec:pattern}
Once we identified the best model for the prediction task, we explore evolution patterns of service community through investigating how do key features affect service community evolution events.
\begin{figure}
    \centering
    \includegraphics[width=\textwidth]{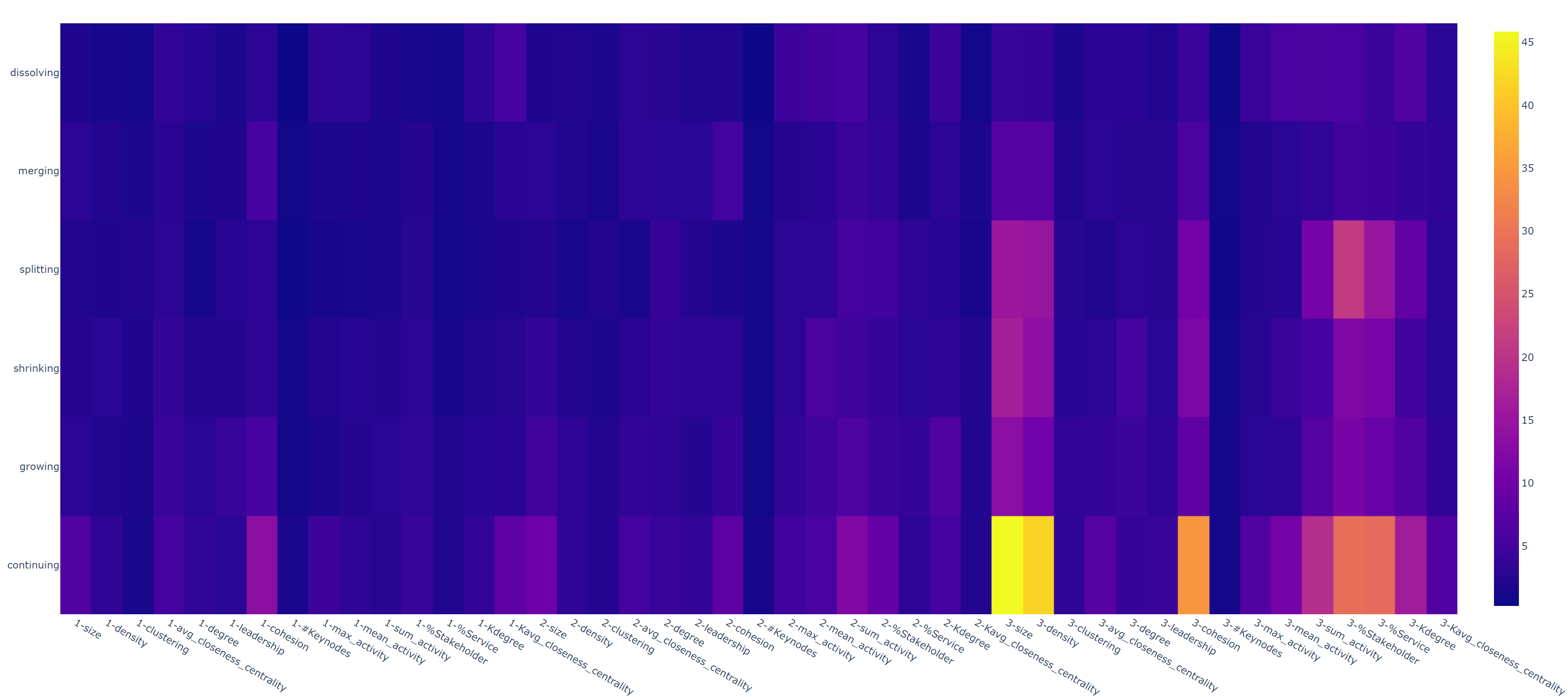}
    \caption{Comparison of important features.}
    \label{fig:heatmap}
        \vspace{-1.5em}

\end{figure}

Fig. \ref{fig:heatmap} shows an heat map of the importance of each selected feature to the prediction of different service community evolution events. It can be observed that for all events, $clustering$ and the number of key nodes in service community contributes little to the prediction. The degree of centralization ($leadership$) of the service community also has little influence on all events, which means that there is no advantage or disadvantage between centralization and decentralization in the field of bike-sharing. For example, both the \textit{Mobike}-centered service community (centralization) and the \textit{HelloBike}-\textit{Youon}-centered service community (decentralization) have a high market share. Although $degree$ and $cohesion$ reflect the degree of service community solidarity from different aspects, $cohesion$ has a greater impact on the events.  It is not surprising that the service community size and the proportion of different types of entities are important features for all events.

Moreover, based on the sensitivity of events to temporal features, we summarize the evolutionary event types as follows:
\begin{enumerate}
    \item \textbf{Full time sensitivity}: for \textit{continuing}, it sensitive to the features of the all time periods, and this sensitivity gradually declines with time.
    \item \textbf{Nearest time sensitivity}: for \textit{growing}, \textit{shrinking} and \textit{splitting}, they are sensitive to the immediate history of the service community rather than more distant history.
    \item \textbf{Not sensitive to time}: for \textit{merging} and \textit{dissolving}, they are not sensitive to time factors, but more sensitive to some specifical features.
\end{enumerate}

\begin{figure}
    \centering
    \includegraphics[width=\textwidth]{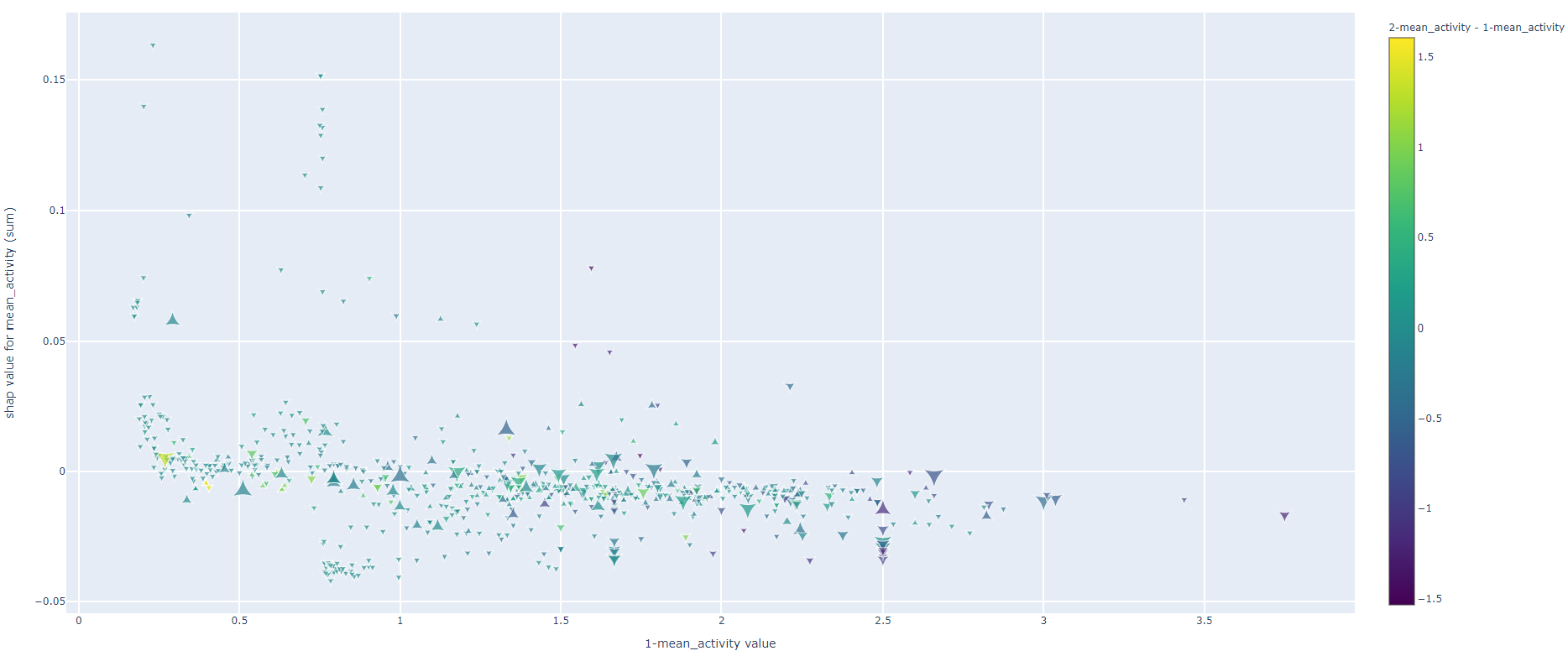}
    \caption{The dependence between $dissolving$ event and $mean\_activity$}
    \label{fig:dissolving_dependency}
        \vspace{-1.5em}
\end{figure}
With SHAP values, we can further understand how a feature affects an evolutionary event in detail. Fig. \ref{fig:dissolving_dependency} shows the dependence between $dissolving$ event and $mean\_activity$. The vertical axis is the sum of SHAP values of $mean\_activity$ of three consecutive service communities, where the positive value indicate that $mean\_activity$ promotes the $dissolving$ and negative value indicate $mean\_activity$ prevents the $dissolving$. The horizontal axis the $mean\_activity$ of the first service community. The color is used to indicate the difference in $mean\_activity$ between the second community and the oldest community. Shape is used to represent the difference between the nearest community, where $\bigtriangleup$ means that the difference is positive, while $\bigtriangledown$ means negative. And the shape size means the absolute value of the difference.

\begin{figure}
    \centering
    \includegraphics[width=\textwidth]{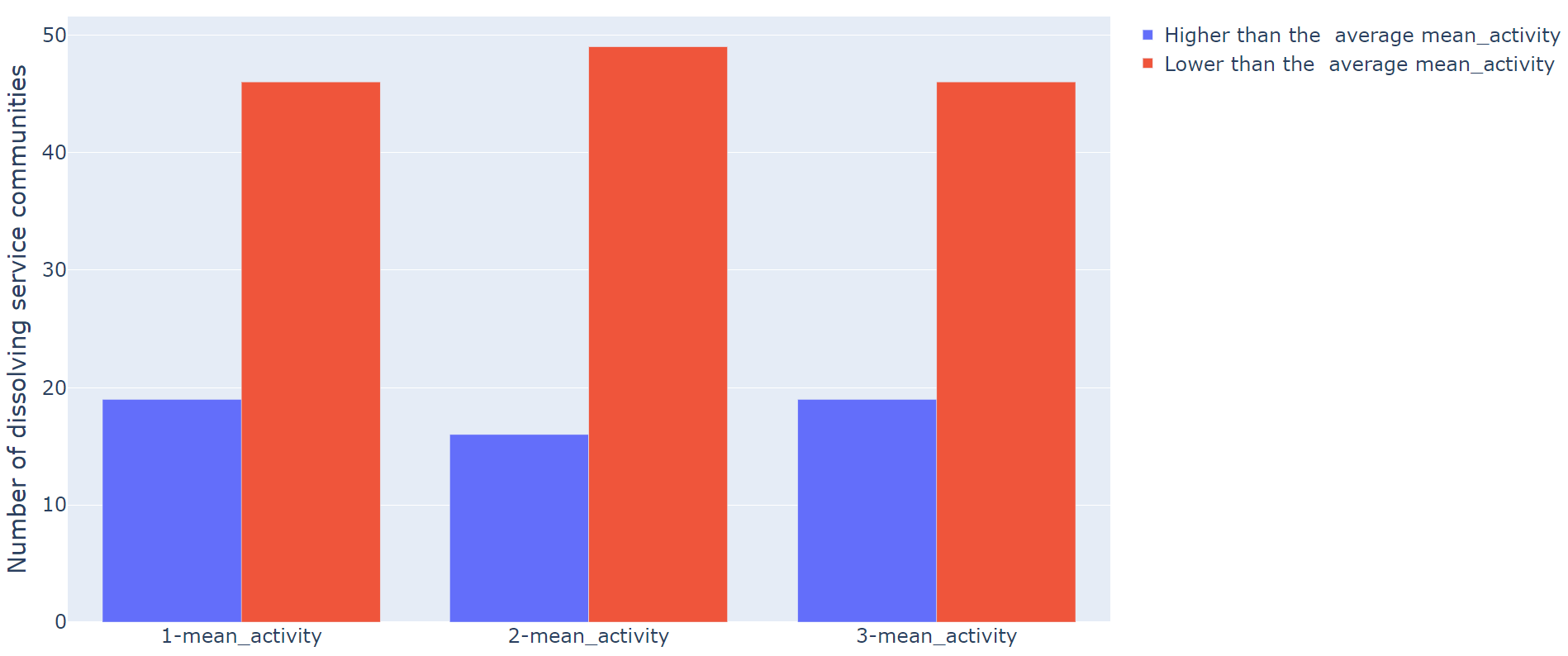}
    \caption{The correlation between the number of $dissolving$ service communities and their $mean\_activity$ value.}
    \label{fig:proof}
        \vspace{-1.5em}
\end{figure}

It is easy to conclude from the figure that service communities with low $mean\_activity$ and declining $mean\_activity$ are more likely to dissolve. Service communities with high $mean\_activity$ are less likely to dissolve in the short time. This pattern is in line with people's common sense, and we also illustrate the correctness of this pattern through the statistical results of our dataset, as shown in Fig. \ref{fig:proof}. It should be noted that this is a simple pattern used as an example. Managers can combine their domain knowledge to discover more complex patterns with the help of these data analysis results.

\subsection{Decision Support}
The prediction model can also assist managers to make decisions by providing the possibility of each evolutionary event and the influence of each feature on this evolution. 

Fig. 8 give an prediction of the evolution trend of \textit{ofo-centered} service community, where $expected\_values$ is the probability of an evolutionary event in the training sample. From the figure, we can see that the probability of \textit{continuing}, \textit{growing}, \textit{shrinking}, \textit{splitting}, \textit{merging}, and \textit{dissolving} occurring in \textit{ofo}-centered service community is $0.0\%$, $4.0\%$, $62.0\%$, $19.7\%$, $8.3\%$ and $6.0\%$ respectively. Therefore, the most likely evolution for \textit{ofo}-centered service community is \textit{shrinking} and \textit{splitting}. Normally, both \textit{shrinking} and \textit{splitting} are negative evolutions, so when a new service provider wants to select partners and enter the bike-sharing market, it is not a good choice for him to join the \textit{ofo}-centered service community, he may need to find those service community that are more likely to grow to join.

\begin{figure}
    \centering
    \includegraphics[width=\textwidth]{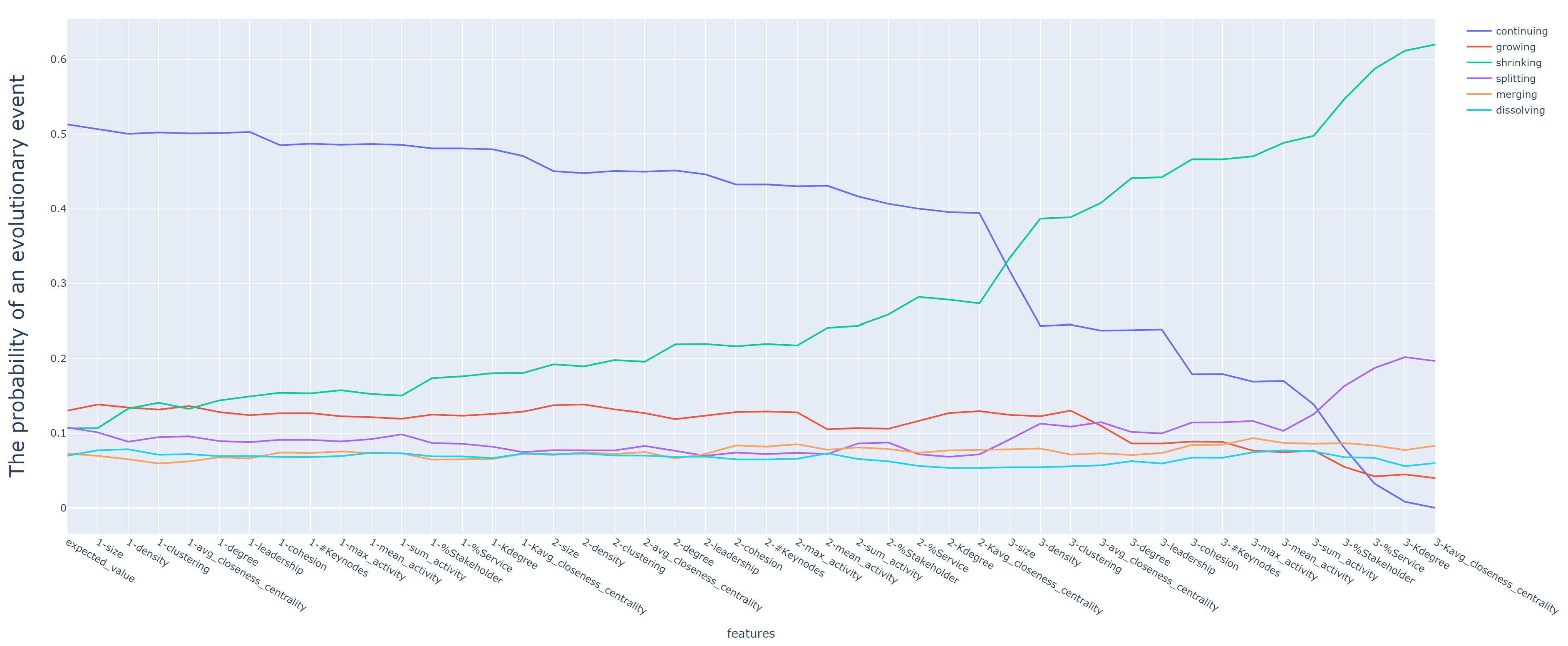}
    \caption{An explainable prediction of the evolution trend of \textit{ofo}-centered service community}
    \label{fig:decision}
    \vspace{-1.5em}
\end{figure}
The figure also shows that the low service community $density$ and the imbalance of the proportion of stakeholders and services in the service community are the main reasons for the \textit{shrinking} and \textit{splitting} of \textit{ofo}-centered service community. Therefore, for stakeholders who are already in the \textit{ofo}-centered service community and want to maintain the stability of the service community, strengthening the cooperation among stakeholders within the community and developing more service functions might be a feasible method. Additionally, the curve of $splitting$ also confirms the pattern summarized in Section \ref{sec:pattern} that $splitting$ is sensitive to the immediate history of the service community.

\vspace{-1em}
\section{Conclusion}
In this paper, we explore the evolution law of service ecosystem by predicting and explaining the evolution of service community. We achieve this by propose a service community evolution analysis and prediction framework. This framework includes service ecosystem snapshot generation, service community detection, service community evolution tracking, service community feature representation, and service community evolution prediction and interpretation. 

We did a case study on the bike-sharing service ecosystem, the result shows that some unique insights on the evolution of service ecosystem can be obtained for the perspective of service community, and the explainable prediction of service community evolution can assist in business decision making.

We are proud that the all the data and tools used in the experiments, have been fully open-sourced. We do hope researchers in this domain use them for more deep research.

\section{Acknowledgment}
Research in this paper is partially supported by the National Key Research and Development 
Program of China (No 2018YFB1402500), the National Science Foundation of China 
(61832004, 61772155, 61802089, 61832014).

\bibliographystyle{splncs04}
\bibliography{reference}

\begin{thebibliography}{10}
\providecommand{\url}[1]{\texttt{#1}}
\providecommand{\urlprefix}{URL }
\providecommand{\doi}[1]{https://doi.org/#1}

\bibitem{10.1007/978-3-030-03596-9_44}
Adeleye, O., Yu, J., Yongchareon, S., Han, Y.: {Constructing and Evaluating an
  Evolving Web-API Network for Service Discovery}. Service-Oriented Computing,
  16th International Conference pp. 603--617 (2018)

\bibitem{blondel2008fast}
Blondel, V.D., Guillaume, J.L., Lambiotte, R., Lefebvre, E.: Fast unfolding of
  communities in large networks. Journal of statistical mechanics: theory and
  experiment  \textbf{2008}(10),  P10008 (2008)

\bibitem{10.1007/s13278-012-0058-8}
Bródka, P., Saganowski, S., Kazienko, P.: {GED: the method for group evolution
  discovery in social networks}. Social Network Analysis and Mining
  \textbf{3}(1),  1--14 (2012)

\bibitem{8491668}
{Dakiche}, N., {Tayeb}, F.B., {Slimani}, Y., {Benatchba}, K.: Sensitive
  analysis of timeframe type and size impact on community evolution prediction.
  In: 2018 IEEE International Conference on Fuzzy Systems (FUZZ-IEEE). pp.~1--8
  (2018)

\bibitem{10.1145/3297280.3297484}
Dakiche, N., Tayeb, F.B.S., Slimani, Y., Benatchba, K.: Community evolution
  prediction in dynamic social networks using community features’ change
  rates. In: Proceedings of the 34th ACM/SIGAPP Symposium on Applied Computing.
  p. 2078–2085. SAC ’19, Association for Computing Machinery, New York, NY,
  USA (2019)

\bibitem{fokaefs2014wsdarwin}
Fokaefs, M., Stroulia, E.: Wsdarwin: Studying the evolution of web service
  systems. In: Advanced Web Services, pp. 199--223. Springer (2014)

\bibitem{10.1109/tase.2013.2297026}
Huang, K., Fan, Y., Tan, W.: {Recommendation in an Evolving Service Ecosystem
  Based on Network Prediction}. IEEE Transactions on Automation Science and
  Engineering  \textbf{11}(3),  906--920 (2014)

\bibitem{10.1007/978-3-662-45391-9_43}
Huang, K., Liu, Y., Nepal, S., Fan, Y., Chen, S., Tan, W.: {A Novel Equitable
  Trustworthy Mechanism for Service Recommendation in the Evolving Service
  Ecosystem} pp. 510--517 (2014)

\bibitem{khan2017network}
Khan, B.S., Niazi, M.A.: Network community detection: A review and visual
  survey. arXiv preprint arXiv:1708.00977  (2017)

\bibitem{liu2020datadriven}
Liu, M., Tu, Z., Xu, X., Wang, Z.: A data-driven approach for constructing
  multilayer network-based service ecosystem models (2020)

\bibitem{NIPS2017_7062}
Lundberg, S.M., Lee, S.I.: A unified approach to interpreting model
  predictions. In: Guyon, I., Luxburg, U.V., Bengio, S., Wallach, H., Fergus,
  R., Vishwanathan, S., Garnett, R. (eds.) Advances in Neural Information
  Processing Systems 30, pp. 4765--4774. Curran Associates, Inc. (2017)

\bibitem{palla2007quantifying}
Palla, G., Barab{\'a}si, A.L., Vicsek, T.: Quantifying social group evolution.
  Nature  \textbf{446}(7136),  664--667 (2007)

\bibitem{saganowski2015predicting}
Saganowski, S., Gliwa, B., Br{\'o}dka, P., Zygmunt, A., Kazienko, P.,
  Ko{\'z}lak, J.: Predicting community evolution in social networks. Entropy
  \textbf{17}(5),  3053--3096 (2015)

\bibitem{8094458}
Sampaio, A.R., Kadiyala, H., Hu, B., Steinbacher, J., Erwin, T., Rosa, N.,
  Beschastnikh, I., Rubin, J.: Supporting microservice evolution. In: 2017 IEEE
  International Conference on Software Maintenance and Evolution (ICSME). pp.
  539--543 (Sep 2017)

\bibitem{wan2016incorporating}
Wan, Y., Chen, L., Yu, Q., Liang, T., Wu, J.: Incorporating heterogeneous
  information for mashup discovery with consistent regularization. In:
  Pacific-Asia Conference on Knowledge Discovery and Data Mining. pp. 436--448.
  Springer (2016)

\bibitem{xu2015big}
Xu, X., Sheng, Q.Z., Zhang, L.J., Fan, Y., Dustdar, S.: From big data to big
  service. Computer (7),  80--83 (2015)

\bibitem{10.1109/icws.2014.17}
Zhong, Y., Fan, Y., Huang, K., Tan, W., Zhang, J.: {Time-Aware Service
  Recommendation for Mashup Creation in an Evolving Service Ecosystem}. 2014
  IEEE International Conference on Web Services pp. 25--32 (2014)

\end{thebibliography}

\end{document}